          [2001/04/25 1.1 (PWD)]
\documentclass[a4paper,twocolumn]{esapub} 
\usepackage{natbib}
\usepackage{epsfig}

\newcommand{\apj}{{\it ApJ, }}

\newcommand{\icar}{{\it Icarus, }}
\newcommand{\mnr}{{\it MNRAS, }}

\newcommand{\ana}{{\it A\&A, }}

\title{Disc--planet, planet--planet, and star--planet interactions during
planet formation}
\author{Richard P. Nelson}
\author{John C.B. Papaloizou}
\affil{Astronomy Unit, Queen Mary, University of London, Mile End Road,
London, E1 4NS, U.K., R.P.Nelson@qmul.ac.uk, J.C.B.Papaloizou@qmul.ac.uk}

\begin{document}

\maketitle

\begin{abstract}
In this article we present results from three on--going projects
related to the formation of protoplanets in protostellar discs.
We present the results of simulations that model
the interaction between embedded protoplanets and disc models undergoing
MHD turbulence. 
We review
the similarities and differences that arise when the disc
is turbulent as opposed to laminar (but viscous), and present the first
results of simulations that examine the tidal interaction between low mass
protoplanets and turbulent discs. \\
We describe the results of simulations of Jovian mass protoplanets
forming in circumbinary discs, and discuss the range of possible
outcomes that arise in hydrodynamic simulations. \\
Finally, we report on
some preliminary simulations of three protoplanets of Jovian mass that
form approximately coevally within a protostellar disc. We describe
the conditions under which such a system can form a 
stable three planet resonance.
\end{abstract}

\section{Introduction}

The on--going discovery of extrasolar planets has reinvigorated efforts
to understand the formation and evolution of planetary systems
(Mayor \& Queloz 1995; Marcy \& Butler 1996; 
Marcy, Cochran, \& Mayor 1999;
Vogt et al. 2002; Santos et al. 2003).
As observations are carried out over longer time scales, 
and with increasing
sensitivity, the physical properties of the observed planetary systems are
set to diversify significantly. At the present time, however,
all of the known planets are Jovian--like gas giants.
For this reason, much of the theoretical research currently underway is
examining the various stages of formation and evolution of giant protoplanets.

The most widely accepted theory of how giant planets form, the so--called
core instability model, suggests that
a multi stage process operates. The solid component of protostellar discs
gradually coagulates to form a solid core of around 15 Earth masses,
onto which a gaseous envelope accretes (e.g. Bodenheimer \& Pollack 1986;
Pollack et al. 1996). An alternative model suggests that giant
protoplanets form {\it via} gravitational instability of a 
protostellar disc (e.g. Boss 2001). In either scenario, interaction between
the forming protoplanet and the protostellar disc is likely to significantly
affect the evolution, leading to orbital migration.

In the standard picture of disc--planet interactions,
a protoplanet exerts torques on a protostellar disc
through the excitation of spiral density waves at Lindblad resonances
(e.g. Goldreich \& Tremaine 1979).
These waves carry  an associated angular momentum flux
which is deposited in the disc
where the
waves are damped. This process results in a negative  torque acting on the
protoplanet from the outer disc and a positive torque
acting on it from the  disc interior to its orbit. For most disc models,
the outer disc torque is dominant, leading to inward migration (Ward 1997).

A sufficiently massive protoplanet can open up an annular gap in a viscous
disc
centred on its orbital radius (Papaloizou \& Lin 1984).
For  typical protostellar disc models the protoplanet
needs to be approximately a Jovian
mass for gap formation to occur.
Recent simulations (Bryden et al. 1999; Kley 1999; Lubow, Seibert, \&
Artymowicz 1999;
D'Angelo, Henning, \& Kley 2002)
examined the formation of gaps by giant protoplanets,
and also estimated the maximal  gas accretion rate onto them.
The orbital evolution
of a Jovian mass protoplanet embedded in a standard laminar
viscous protostellar
disc model
was studied by Nelson et al. (2000). They found that
that gap formation and accretion of the inner disc by the central mass
led to the formation of a low density inner cavity in which the planet
orbits.  Interaction with the outer disc  resulted in inward type II
migration on a time scale of a few $\times 10^5$ yr. Gas accretion during 
migration allows the protoplanet to grow to $\simeq 3$--4 Jupiter masses.
The disc models in these
studies all  adopted an anomalous
disc viscosity   modelled through the Navier--Stokes equations
without consideration of its origin.

The most likely origin of the viscosity is through
MHD turbulence resulting from the magnetorotational
instability (MRI) (Balbus \& Hawley 1991)
and it has recently become possible through
improvements in computational resources
to  simulate  discs in which this  underlying mechanism responsible
for angular momentum transport is explicitly calculated.
This is necessary because the turbulent fluctuations
do not necessarily result in transport phenomena that can be
modelled with the Navier--Stokes equation.

To this end Papaloizou \& Nelson (2003) and Nelson
\& Papaloizou (2003a) developed models of turbulent
protostellar accretion discs and considered the interaction
with a giant protoplanet of $5$ Jupiter masses.
The large mass was chosen to increase the scale of the interaction
so reducing the computational resources required.
This protoplanet was massive enough to maintain
a deep gap separating the inner and outer disc and
exert torques characteristic of type II migration.
More recent work has expanded the range of protoplanet masses
examined (Papaloizou, Nelson, \& Snellgrove 2003; Nelson \& Papaloizou 2003b).
We discuss some of the main points of this work in later sections.

The majority of
the planets so far detected orbit around single solar--type stars,
but there have been
also been detections
in binary systems
[e.g. $\gamma$ Cephei (Hatzes et al. 2003),
16 Cygni B (Cochran et al. 1997)].
Most field stars appear to be members
of binary systems (Duquennoy \& Mayor 1991).
For the longer period systems planets may 
orbit around
one member of the binary. In the shorter period systems they could orbit
stably around both stars (i.e. circumbinary planets).

The majority of T Tauri stars, whose discs are thought to be the sites
of planet formation, also appear to be in binary or multiple systems,
(Ghez, Neugebauer \& Matthews 1993;
Leinert et al. 1993; Mathieu et al. 2000).
Most
have sufficiently large separations that it is expected that each component
will have its own circumstellar disc. For shorter period systems, however,
one expects the existence of a circumbinary disc, a number of which have
been observed
(e.g. DQ Tau, AK Sco, UZ Tau, GW Ori, GG Tau).

The confirmed existence of planets in binary systems,
combined with the fact that
binary systems appear to be common, and to be present during the T Tauri phase,
means that it is of interest to explore
how stellar multiplicity affects planet formation, and
post-formation planetary orbital evolution, including formation in circumbinary
discs.
Previous work examined the stability of planetary orbits
in binary systems using N--body simulations
(Dvorak 1986; Holman \& Wiegert 1999). This work showed
that there is a critical ratio of planetary to binary semimajor
axis for stability, depending on the binary mass ratio, $q_{bin}$, and
eccentricity $e_{bin}$.
A recent paper (Quintana et al. 2002) explored the late stages
of terrestrial planet formation
in the $\alpha$ Centauri system. This work concluded that the
binary companion can help speed up
planetary accumumlation by stirring up the planetary embryos, thus increasing
the collision rate.
\begin{figure}
\centerline{
\epsfig{file=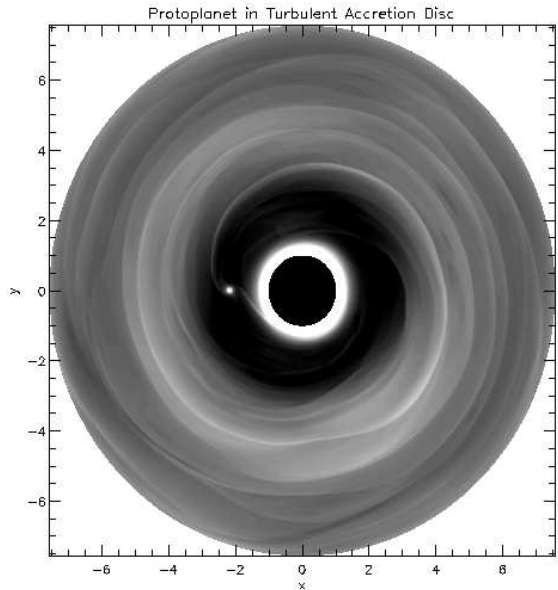,width=8cm} }
\caption[]
{This figure shows midplanet density distribution of 5 Jupiter mass protoplanet
in a turbulent disc}
\label{fig1}
\end{figure}
Recent work  by Kley \& Burkert (2000)
examined the effect that an external binary
companion can have on the migration and mass accretion of a giant protoplanet
forming in a circumstellar disc. They found that for sufficiently close
companions, both the mass accretion rate and the orbital migration
rate could be increased above that expected for protoplanets forming
around single stars.

The question of how protoplanet evolution is affected in circumbinary
discs has been examined recently by Nelson (2003a). This work
explored the evolution
of Jovian mass protoplanets forming in circumbinary discs using
hydrodynamic simulations of a binary star plus protoplanet system interacting
with a viscous protostellar disc. The models apply primarily to
binaries with orbital
periods of $\sim 1$ yr, and semimajor axes of $\sim 1$ AU that have
protoplanets
forming at a radius of a few AU in the circumbinary disc,
although the results can be scaled to apply to different parameters.
It is well known that a giant protoplanet embedded in a disc
around a single
star undergoes inward migration driven by the viscous evolution of the
disc (e.g. Nelson et al. 2000).
The work by Nelson (2003a) examines how this process is affected
when the central star is replaced by a close binary system, and delineates
the various modes of behaviour that arise depending on the properties of the
system (e.g. binary mass ratio $q_{bin}$, binary eccentricity $e_{bin}$, etc).
We present some of the main results of this work in later sections.

\begin{figure}
\centerline{
\epsfig{file=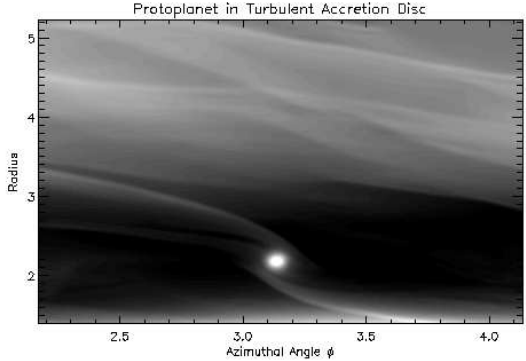,width=7cm} }
\caption[]
{This figure shows a close-up of the midplane density distribution for a
5 Jupiter mass protoplanet
in a turbulent disc}
\label{fig2}
\end{figure}

A number of the known extrasolar planets exist in multi--planet systems.
Three of these systems contain a pair of planets that are in
mean motion resonance (GJ876, HD82943, 55 Cancri).
The most likely explanation for these resonant systems
is approximately coeval formation of the planet pair, followed by disc--induced
differential migration leading to resonant capture (e.g. Snellgrove, Papaloizou,
\& Nelson 2001; Lee \& Peale 2002). In later sections 
we present some preliminary results that
examine the plausibility of three--planet resonances being discovered
in which the three planets are of approximately Jovian mass.
The Jovian satellite system displays such a configuration with
Io, Europa, and Ganymede all participating in the Laplace resonance
(e.g Peale, Cassen, \& Reynolds 1979). Here Io and Europa are in 2:1
resonance, and Europa and Ganymede are simultaneously in 2:1 resonance,
leading to a 4:2:1 relationship between the mean motion of Io, Europa, and
Ganymede.
Our preliminary calculations indicate that three planet resonances may indeed 
be established, but that the 4:2:1 relation is unstable.
However, a situation in which one of the planet pair is in a 3:1 resonance,
leading to a 6:2:1 or 6:3:1 relation between the mean motions, appears to be
stable, suggesting that such configuration may be found among the population
of extrasolar planets.

This article is organised as follows. In section~\ref{Turb} we present the results of simulation of high and low mass protoplanets interacting with turbulent
accretion discs. In section~\ref{binary} we describe the results of
simulations that examine the evolution of giant protoplanets forming in
circumbinary accretion discs. In section~\ref{3planet} we present some
preliminary results of three planet systems leading to the formation of
three-planet resonances. Finally we summarise the results presented in this
article in section~\ref{summary}.

\section{Turbulent Disc -- Protoplanet Interactions}
\label{Turb}
Most of the previous work examining the interaction between
protostellar disc models and embedded protoplanets have used
the Navier--Stokes equations to simulate laminar disc models 
with an anomalous viscosity coefficient (e.g. Bryden et al. 1999; Kley 1999;
Nelson et al. 2000). The most likely source of disc viscosity, however,
is MHD turbulence that arises because of the MRI (Balbus \& Hawley 1991).
We present simulations of magnetic, turbulent discs interacting with
protoplanets of different mass in the following sections.

\subsection{Giant Protoplanets}
\label{gplanet}
\begin{figure}
\centerline{
\epsfig{file=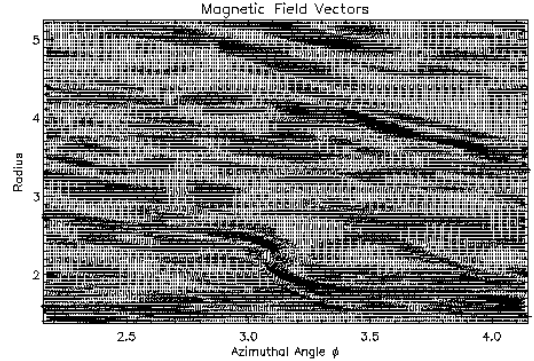,width=7cm} }
\caption[]
{This figure shows magnetic field vectors for equivalent region shown in fig.2.
}
\label{fig3}
\end{figure}
A detailed discussion of the results presented in this section may be found
in Nelson \& Papaloizou (2003a). The underlying disc model used is described
in detail in Papaloizou \& Nelson (2003). The disc was a cylindrical disc
model with a locally isothermal equation of state such that the effective
aspect ratio $H/R=0.1$ throughout. The inital disc model was initiated
with a zero--net flux vertical magnetic field that varied sinusoidally
with radius, and was allowed to run until a statistically steady state
turbulent disc model was obtained. The initial value of the volume averaged
plasma $\beta$ was $< \beta > \simeq 1000$, where $\beta$ is the ratio of the
thermal gas pressure to the magnetic pressure.
The final turbulent state generated a
volume averaged value of the Shakura--Sunyaev 
stress parameter $<\alpha> \simeq 5 \times 10^{-3}$.
A protoplanet of mass $M_p=5$ Jupiter masses was placed in the disc
at a radius $r_p=2.2$. Prior to placing the protoplanet in the disc, a
partial gap was made in the disc in order to avoid the generation of
transient features in the flow associated with the gap opening
process.

A snapshot of the final state of the disc with the protoplanet is shown
in figure~\ref{fig1}, after $\sim 100$ planetary orbits.
The protoplanet showed a clear tendency towards gap formation by
deepening and widening the initially imposed partial gap during the
simulation. Broadly speaking, the visual appearance of the disc
in figure~\ref{fig1} differs significantly from that obtained in an equivalent
simulation with a laminar, viscous disc (e.g. Nelson et al. 2000; Nelson \&
Papaloizou 2003a), with the turbulent disc showing a less regular 
and more time dependent structure,
in which the spiral waves have a more diffused appearence.

In the region of the wakes the disc and its turbulence are strongly perturbed.
Magnetic field vectors in the disc midplane in the neighbourhood
of the planet are illustrated
in figure~\ref{fig3} along with a corresponding 
density plot for the equivalent region in figure~\ref{fig2}.
An inspection of the magnetic field vectors indicates that these
tend to line up along the location of the wakes but in a somewhat
broadened region slightly behind the shocks.
In this way an ordered structure appears to be imposed
on the flow and magnetic field by the protoplanet.
The magnetic stress is largely communicated in these ordered regions,
leading to a significant change in the magnetic contribution to the
stress parameter $\alpha$ there (Nelson \& Papaloizou 2003a).

\begin{figure}
\centerline{
\epsfig{file=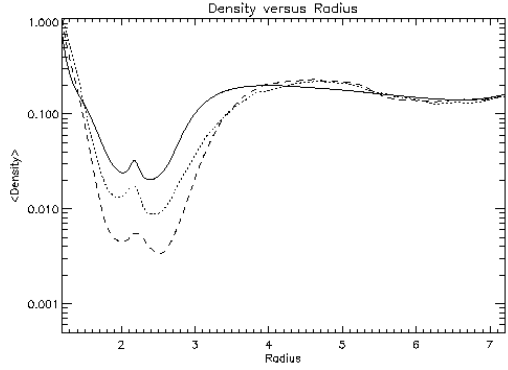,width=7cm} }
\caption[]
{This figure shows the azimuthally averaged
radial density distribution for MHD turbulent disc (dotted line),
and laminar disc with (solid line) and without (dashed line)
alpha viscosity}
\label{fig4}
\end{figure}

In addition to running the full MHD turbulent disc model, we have also
run some 2D laminar $\alpha$ disc models for the purposes of comparison.
Two 2D model were run with $\alpha=0$ and $\alpha=5 \times 10^{-3}$, 
respectively.
For initial conditions in these 2D models, we took the midplane
density distribution of the 3D turbulent model just prior to the
addition of the planet, switched off the magnetic field, and introduced
the required value of $\alpha$ in the Navier--Stokes viscosity. The
implementation of the Navier--Stokes viscosity is described in
Nelson et al. (2000).

\begin{figure*}
\centerline{
\epsfig{file=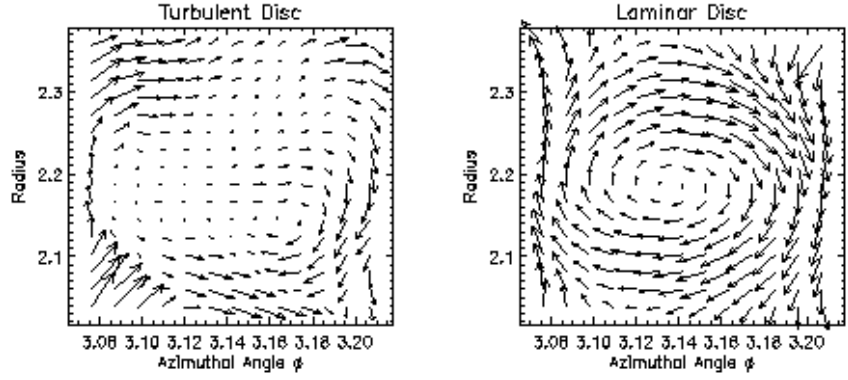,width=12cm}}
\caption[]
{This figure shows velocity vectors for material in the planet Hill sphere.
The left hand panel shows the MHD run, and the right hand panel a laminar
disc run.}
\label{fig5}
\end{figure*}

The azimuthally averaged surface density distribution
for these models, plus the midplane density distribution for the
turbulent model, is plotted in figure~\ref{fig4}. Each of the models have been
run for a total time of $\simeq 2050$ time units, corresponding to $\sim 100$
planetary orbits. It is clear that the $\alpha=0.0$ 2D model (dashed line)
has the deepest
and widest gap, as expected. It is also apparent that the turbulent
model (dotted line) has a deeper and wider gap than the 2D $\alpha=5 \times 10^{-3}$ model,
even though
a volume averaged estimate for the underlying turbulent disc model yields
an effective $\alpha \simeq 5 \times 10^{-3}$ (see figure 18 in Papaloizou \& 
Nelson 2003).
Thus, the turbulent model behaves as if it has a somewhat
smaller $\alpha$ than
reasonable estimates suggest it has.
This may arise for the following reasons.
First, a Navier--Stokes viscosity with anomalous
viscosity coefficient provides a source of constantly acting friction
in the disc, such that it can induce a steady mass flow into the
gap region. The turbulence, however, does not operate as a constant
source of friction that generates steady inflow velocities. Instead
it generates large velocity fluctuations that may be much
larger than the underlying inflow velocity arising from
the associated angular momentum transport. Results presented in Papaloizou \&
Nelson (2003)
indicate that a process of time averaging the turbulent velocity
field is required over long time periods before these fluctuations
can be averaged out to reveal the underlying mass flow. The disc material
in the vicinity of the planet experiences periodic high amplitude perturbations
induced by the planet
on a time scale much shorter than the required averaging time scale,
so that the disc response is expected to differ from that in the case of a
disc with Navier--Stokes viscosity.
A second
plausible reason for the apparently lower $\alpha$ is
that the existence of the magnetic field in the turbulent disc
allows for field lines to connect across the gap region, and to enable
angular momentum transport across the gap. In this way the magnetic
field actually helps the planet to maintain the gap (Nelson \& Papaloizou 2003a).
The differences in gap structure found here
suggest that the accretion rate
onto a protoplanet in an MHD turbulent disc is likely to be less
than previous estimates based on laminar  $\alpha$ disc models indicate
[e.g. Bryden et al. (1999); Kley (1999); Lubow, Seibert,
\& Artymowicz (1999); Nelson et al. (2000); D'Angelo, Henning, \& Kley (2002)],
but not by a large factor.

The velocity field within the Hill sphere of the planet is shown in the left
hand panel of figure~\ref{fig5} for the turbulent disc. An equivalent
plot for a laminar disc is shown in the right hand panel of this figure.
The material that enters the Hill sphere of the protoplanet and becomes
gravitationally bound to it is normally expected to circulate around it
by virtue of its angular momentum. In the calculation presented
here the protoplanet does not accrete gas, and is modelled as
a softened point mass. Consequently it is expected that material that
enters the Hill sphere will form a hydrostatic `atmosphere' around the
planet that is supported through pressure and rotation. In figure~\ref{fig5}
we would expect to see circulation occurring in a clockwise fashion for
both the MHD turbulent disc and the laminar disc.
However, it is apparent that the circulating pattern has been disrupted
in the magnetised disc, indicating
that magnetic breaking may have been responsible for this (Nelson \& Papaloizou
2003a).
The expected circulation is observed in the non magnetic run performed
at the same numerical resolution as shown in the right hand panel of
figure~\ref{fig5}. Magnetic braking of material that forms a circumplantary
disc inside the Hill sphere will have significant implications for the mass 
accretion onto the protoplanet.

\subsection{Low mass protoplanets}
In this section we present the results of simulations that examine the
interaction of turbulent discs with low mass, embedded protoplanets.
A detailed discussion of the results presented in this 
section is given in Papaloizou, Nelson \& Snellgrove (2003)
and Nelson \& Papaloizou (2003b).
The underlying disc model used in the simulations with low mass
protoplanets differed from that described in section~\ref{gplanet}.
The disc was a cylindrical disc model with a locally isothermal equation
of state. The effective aspect ratio took a constant value $H/r=0.07$.
The disc model was initiated with a zero--net flux toroidal magnetic field
that varied sinusoidally with radius. The initial value of the volume
averaged plasma $\beta$ parameter $<\beta> \simeq 30$. 
The disc model was evolved until a statistically steady state turbulent disc
was obtained. The final turbulent state had an associated volume averaged
stress parameter $<\alpha> \simeq 7 \times 10^{-3}$ (Papaloizou, Nelson, \&
Snellgrove 2003).

\begin{figure}
\centerline{
\epsfig{file=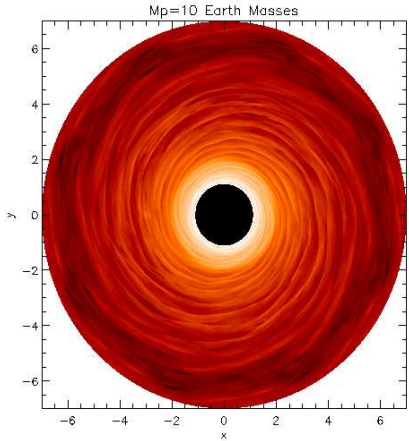,width=6cm} }
\caption[]
{This figure shows a snapshot of a turbulent disc with an embedded
$q=3 \times 10^{-5}$ protoplanet located at ($x$,$y$)=(-3,0).}
\label{fig6}
\end{figure}

Once the final turbulent state had been obtained, low mass protoplanets
were inserted into the disc model. These had mass ratios $q=m_p/M_*=10^{-5}$,
$3 \times 10^{-5}$, and $10^{-4}$, respectively. The gravitational potential
of the protoplanets was softened using a softening parameter $b=0.3H_p$,
where $H_p$ is the disc semi--thickness at the position of the protoplanet.
The primary aim of these simulations was to examine the tidal torques exerted
on non gap forming protoplanets by turbulent accretion discs to estimate
the corresponding migration rates.

\begin{figure}
\centerline{
\epsfig{file=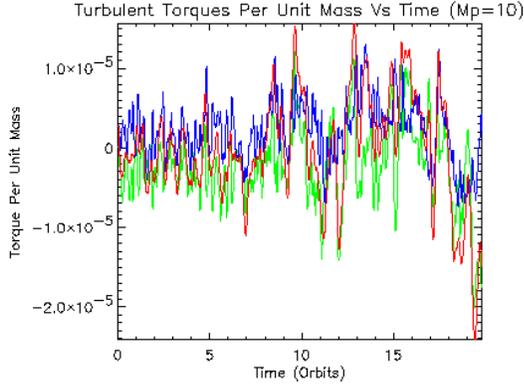,width=7cm} }
\caption[]
{This figure shows the time evolution of the torque per unit mass
exerted on $q=3 \times 10^{-5}$ protoplanet. The torque is clearly dominated
by strong fluctuation due to the turbulence.}
\label{fig7}
\end{figure}

\begin{figure}
\centerline{
\epsfig{file=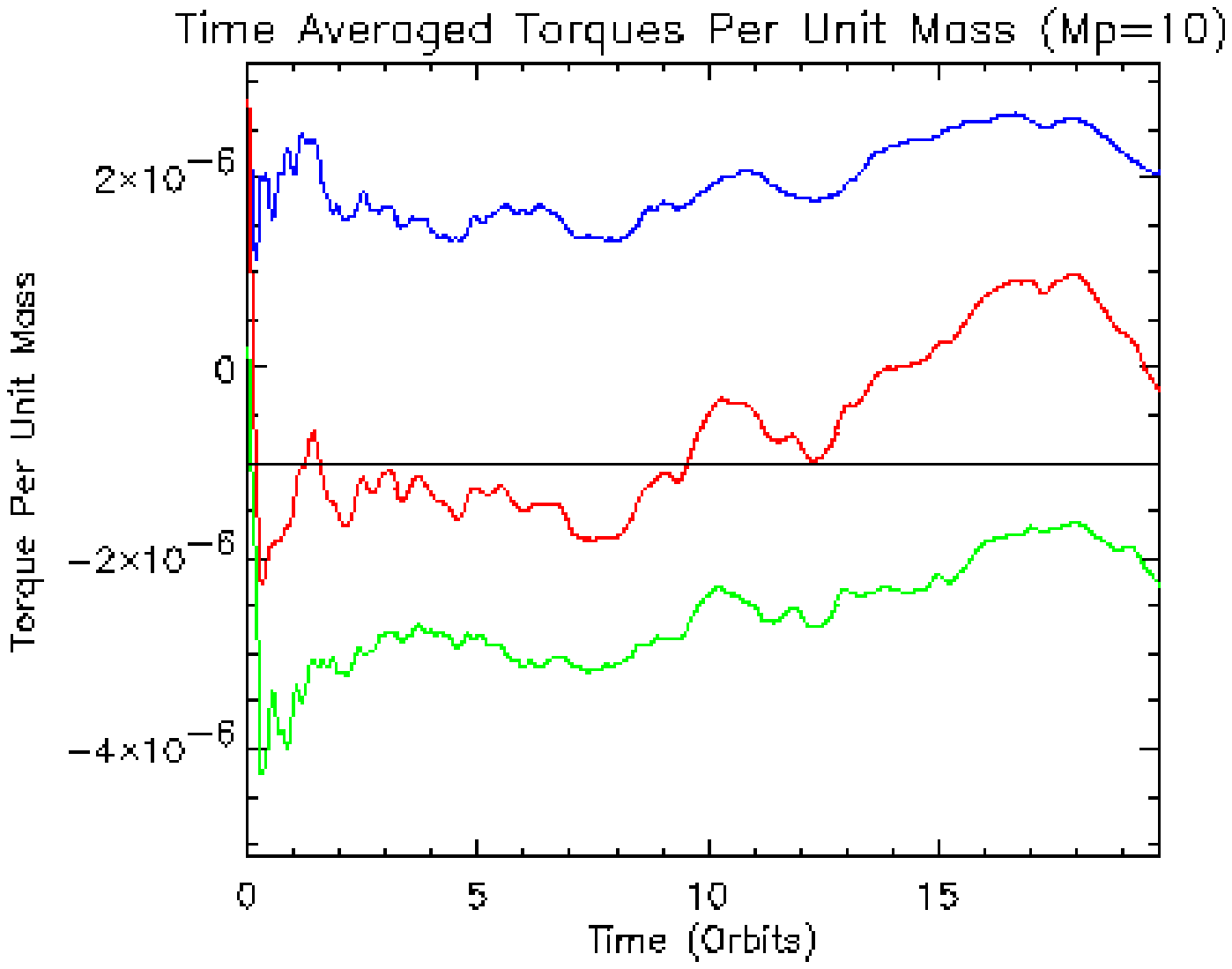,width=7cm} }
\caption[]
{This figure shows the time evolution of the running time average of the
torque per unit mass. The upper line corresponds to the inner disc torque,
the lower line gives the outer disc torque, and the middle line the 
running time average of the total torque. The straight line indicates the
torque obtained in an equivalent laminar disc run.}
\label{fig8}
\end{figure}

A snapshot of a simulation with $q=3 \times 10^{-5}$ is shown in 
figure~\ref{fig6}. This corresponds to a protoplanet of
$m_p \simeq 10$ Earth masses orbiting a solar mass star.
The protoplanet is located at $(x,y)=(-3,0)$, and
is only just visible in this figure since the density
fluctuations generated by the turbulence are in fact of higher
amplitude than the spiral wakes generated by the protoplanet.
The torque per unit mass exerted on the protoplanet was calculated
as a function of time for all simulations, and is shown in figure~\ref{fig7}
for the case with $q=3 \times 10^{-5}$. This figure clearly shows that
the force exerted on the protoplanet by the turbulent disc is dominated by
high amplitude fluctuations. The total torque exerted on the protoplanet is
seen to oscillate between positive and negative values, suggesting that
the migration in this case is likely to occur as a `random walk', rather than
as a monotonic inward drift normally associated with type I migration
(Ward 1997). Figure~\ref{fig8} shows the time
evolution of the running time average of the torque per unit mass.
The straight line shows the torque obtained from an equivalent simulation
with a laminar disc model. The upper line shows the running time average
of the torque due to the inner disc, the lower line the torque due to the outer
disc, and the middle line the running average of the total torque.
The running mean of the total torque is apparently not converging to a well
defined value on the time scale of the simulation, and for a large part of
the simulation indicates that the protoplanet would migrate
{\em outwards} on average. Treating the turbulent
fluctuations as having a Gaussian distribution, we can estimate the time
for the running mean to converge as a function of the amplitude of the
fluctuations. Such an estimate yields a time scale of $\simeq 100$ 
planetary orbits for convergence, longer than we are currently able
to evolve the simulation (Nelson \& Papaloizou 2003b).

The simulations with $q=10^{-5}$ and $q=10^{-4}$ showed qualitatively similar
results to those described above. Although we are unable to run simulations 
for sufficient length to definitively calculate the migration times of
protoplanets in turbulent discs, it is clear that the picture of migration
that emerges differs significantly from that obtained in laminar disc
models. These results offer the possibility that the rapid migration rates
of giant protoplanet cores may in fact be slowed down or stopped by
interaction with the density fluctuations in a turbulent disc.

\section{Giant Planets Forming in Circumbinary Discs}
\label{binary}
In this section we present the results of simulations
that examine the evolution of giant protoplanets that form in
circumbinary discs. A fuller discussion of this work is presented in 
Nelson (2003a).
We consider the interaction between a coplanar
binary and protoplanet
system and a two--dimensional, gaseous, viscous, circumbinary disc
within which it is supposed
the giant protoplanet formed. We do not consider the early evolution of
the protoplanet in this work or address the formation process itself,
but make the assumption that a Jovian mass protoplanet is able to form
and examine the dynamical consequences of this.
\begin{figure}
\centerline{
\epsfig{file=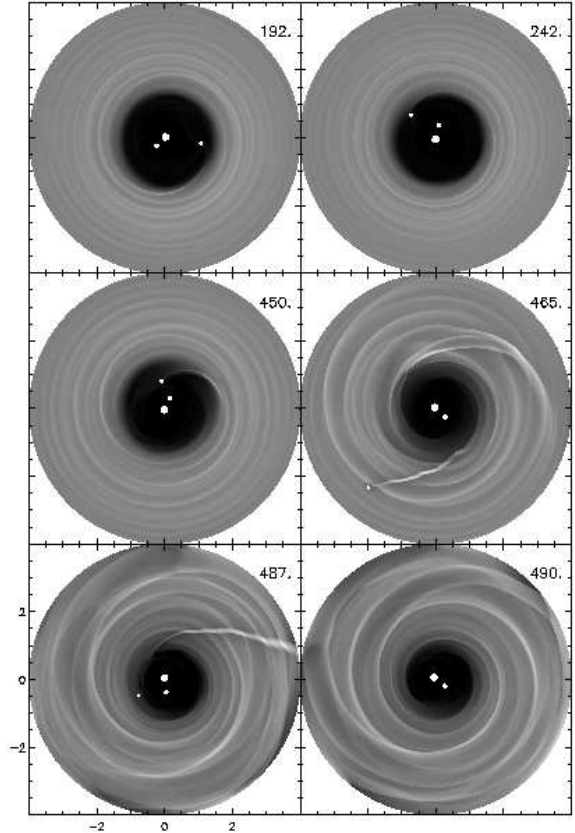,width=8cm} }
\caption[]
{This figure shows surface density contours for run in which the planet is ejected by the binary}
\label{fig9}
\end{figure}

Each of the stellar components and the protoplanet experience
the gravitational force of the other two, as well as
that due to the disc. The disc is evolved using the hydrodynamics
code NIRVANA
(Ziegler \& Yorke 1997). The planet and binary orbits are
evolved using a fifth--order Runge--Kutta scheme (Press et al. 1992).
The force of the planet on the disc, and of the disc on the planet, is softened
using a gravitational softening parameter $b=0.5a_p(H/r)$, where $a_p$ is the
semimajor axis of the planet, and $H/r$ is the disc aspect ratio.
We assume that the mass of the protoplanet is fixed, and so do not allow
accretion of matter from the disc onto the protoplanet.

We adopt a disc model in which the effective aspect ratio $H/r=0.05$, and the
Shakura--Sunyaev viscosity parameter $\alpha=5 \times 10^{-3}$
(Shakura \& Sunyaev 1973).
The surface density  $\Sigma$ is initialised to have an inner cavity within which the
planet and binary orbit (Nelson 2003a).
Simulations
initiated with no inner cavity
show that one is formed by the action of the binary system and planet
clearing gaps in their local neighbourhood. As the planet migrates
in towards the
central binary these gaps join to form a single cavity.
The disc mass is normalised through the choice of $\Sigma_0$ such that
a standard disc model with $\Sigma(r)=\Sigma_0r^{-1/2}$
throughout would contain about 4
Jupiter masses interior to the initial planet radius $r_p$ (assumed in physical
units to be 5 AU). Thus the disc mass interior to the initial planet radius
would be about twice that of a minimum mass solar nebula model.

The total mass of the binary plus protoplanet system is assumed to be 1 M$_{\odot}$.
Dimensionless units are used such that the total mass of the binary system
plus planet $M_{tot}=1$ and the gravitational constant $G=1$.
The initial binary semimajor axis is $a_{bin}=0.4$
in all simulations, and the initial planet semimajor axis $a_p=1.4$.
The simulations are initiated with the binary system having an initial
eccentricity, $e_{bin}$, and the protoplanet is initially in circular
orbit. The binary mass ratio $q_{bin}=0.1$ for all simulations presented
in this section, but larger values were considered in Nelson (2003a).
The unit of time quoted in the discussion of the simulation results
below is the orbital period at $R=1$.

\subsection{Numerical Results}
\begin{figure}
\centerline{
\epsfig{file=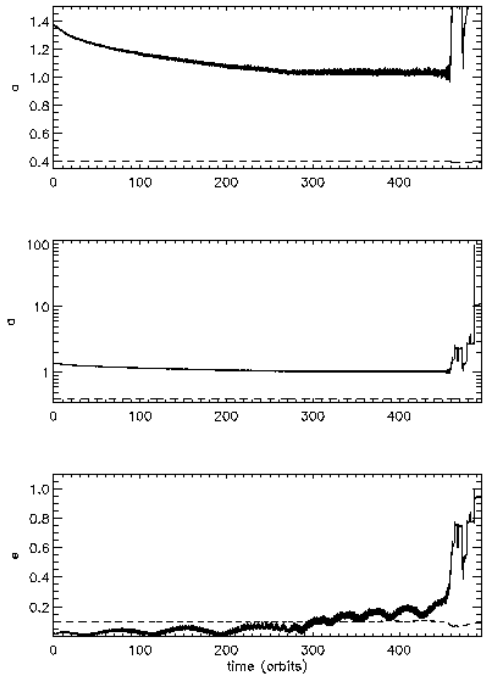,width=7cm} }
\caption[]
{This figure shows semimajor axes and eccentricities for run in which
planet is scattered by the binary.}
\label{fig10}
\end{figure}
The results of the simulations 
can be divided into three categories (Mode 1, Mode 2, and Mode 3), which are
described below, and are
most strongly correlated with
changes in the binary mass ratio, $q_{bin}$,  and
binary eccentricity $e_{bin}$. Changes to the disc mass and/or protoplanet
mass appear to be less important. In some runs the planet enters the
4:1 mean motion resonance with the binary.
The associated resonant angles in the coplanar case are defined by:
\begin{eqnarray}
\psi_1 =  4 \lambda_s - \lambda_p - 3 \omega_s \;\;\;\;\;\;\;\; &&
\psi_2  = 4 \lambda_s - \lambda_p - 3 \omega_p  \\
\psi_3 =  4 \lambda_s - \lambda_p - 2 \omega_s - \omega_p &&
\psi_4  =  4 \lambda_s - \lambda_p - 2 \omega_p -\omega_s \nonumber
\label{res_ang}
\end{eqnarray}
where $\lambda_s$, $\lambda_p$ are the mean longitudes of the secondary star
and protoplanet, respectively, and $\omega_s$, $\omega_p$ are the longitudes
of pericentre of the secondary and protoplanet, respectively. When in
resonance $\psi_3$ or $\psi_4$ should librate,
or all the angles should librate.
In principle the protoplanet is able to enter higher order resonances
than 4:1, such as 5:1 or 6:1, since its initial location lies
beyond these resonance locations. However, none of the simulations presented
here resulted in such a capture. Test calculations
indicate that capture into higher order resonances requires slower planetary
migration rates than those that arise in these simulations. For significantly
faster migration rates the planet may pass through the 4:1 resonance
(Nelson 2003a).

\subsubsection{Mode 1 -- Planetary Scattering}
\label{sec:mode1}
A number of simulations resulted
in a close encounter between the protoplanet and binary system, leading to
gravitational scattering of the protoplanet to larger radii, or into an
unbound state. We label this mode of evolution as `Mode 1'.
Typically the initial scattering
causes the eccentricity of the planet to grow to values
$e_p \simeq 0.9$,
and the semimajor axis to increase to $a_p \simeq 6$ -- 8. In runs that
were continued for significant times after this initial scattering,
ejection of the planet could occur after subsequent
close encounters.

\begin{figure}
\centerline{
\epsfig{file=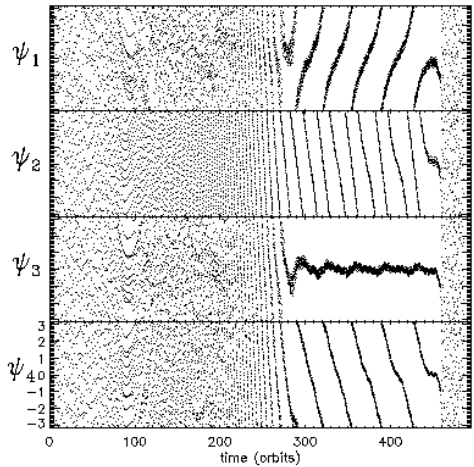,width=7cm} }
\caption[]
{This figure shows the evolution of the resonant anglesd defined in 
equation~\ref{res_ang}}
\label{fig11}
\end{figure}

We will illustrate this mode of evolution using a simulation with
$m_p= 3$ Jupiter masses and $q{bin}=0.1$. A series of snapshots of
the simulation are shown in figure~\ref{fig9}.
Mode 1 evolution proceeds as follows.
The protoplanet migrates in towards the
central binary due to interaction with the circumbinary disc, and
temporarily enters the 4:1 mean motion resonance with the binary. The migration
and eccentricity evolution is shown in figure~\ref{fig10}, and the resonance angles
are shown in figure~\ref{fig11}.
The
resonant angle $\psi_3$ librates with
low amplitude, indicating that the protoplanet is strongly locked into the
resonance.
The resonance drives the eccentricity of the protoplanet upward, until
the protoplanet has a close encounter with the secondary star during or close 
to periapse,
and is scattered out of the resonance into a high eccentricity
orbit with significantly larger semimajor axis. We note that being in resonance
normally helps maintain the stability of two objects orbiting about a central
mass. However, when one of the objects is a star, the large
perturbations experienced by the planet can cause the resonance to break
when the eccentricities are significant. Once out of resonance, the chances
of a close encounter and subsequent scattering are greatly increased.
This provides a method of forming `free--floating planets'.

\subsubsection{Mode 2 -- Near--resonant Protoplanet}
\label{sec:mode2}

\begin{figure}
\centerline{
\epsfig{file=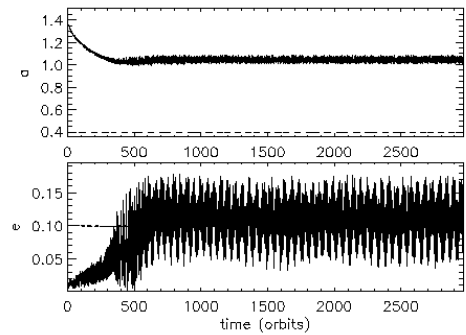,width=7cm} }
\caption[]
{This figure shows semimajor axes and eccentricities for the Mode 2 run 
described in the text.}
\label{fig12}
\end{figure}

A mode of evolution was found in some of the simulations 
leading
to the protoplanet orbiting stably just outside of the 4:1
resonance. We label this mode of evolution as `Mode 2'.
Mode 2 evolution is illustrated by a simulation for which $m_p=1$,
$q_{bin}=0.1$, and $e_{bin}=0.1$. The evolution of the orbital elements
are shown in figure~\ref{fig12}.
Here, the protoplanet migrates inwards and
becomes {\em weakly} locked into the 4:1 resonance,
with the resonant angle $\psi_3$ librating with large
amplitude.
The resonance becomes undefined and
breaks when
$e_p=0$ momentarily during the high amplitude oscillations of $e_p$
that accompany the high amplitude librations of $\psi_3$.
The protoplanet
undergoes a period
of outward migration through interaction with the disc
by virtue of the eccentricity having reattained values of $e_p \simeq 0.17$
once the resonance is broken.
Calculations by Nelson (2003b) show that gap--forming protoplanets
orbiting in tidally truncated discs undergo outward migration if they
are given eccentricities of this magnitude impulsively, due to the sign of the
torque exerted by the disc reversing for large eccentricities.
The outward migration moves the planet to a safer
distance away from the binary, helping to avoid instability.  \\
Once the protoplanet has migrated to just beyond the 4:1 resonance the
outward migration halts, since its eccentricity reduces
slightly, and the planet
remains there for the duration of the simulation. The system achieves a balance
between eccentricity damping by the disc and eccentricity excitation by the binary,
maintaining a mean value of $e_p \simeq 0.12$ (Nelson 2003a).
The torque exerted by the disc
on the protoplanet
is significantly weakened by virtue of the finite eccentricity (Nelson 2003b),
preventing the planet from migrating back towards the binary. \\
Continuation of this run in the absence of the
disc indicates that the planet remains stable for over $6 \times 10^6$ orbits.
This is in good agreement with the stability criteria obtained by
Holman \& Wiegert (1999) since the protoplanet lies just outside of the zone of
instability found by their study.

\subsubsection{Mode 3 -- Eccentric Discs}
\label{sec:mode3}
\begin{figure}
\centerline{
\epsfig{file=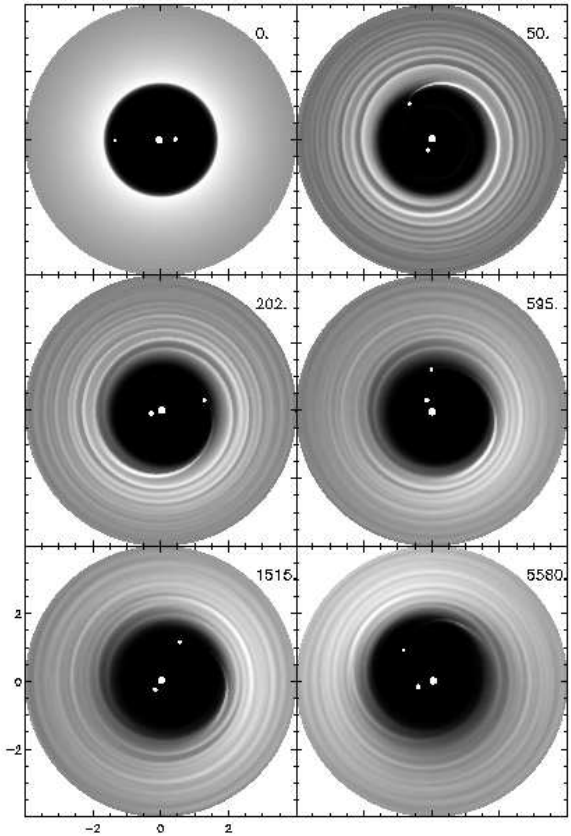,width=8cm} }
\caption[]
{This figure shows contours of surface density for the Mode 3 run described
in the text.}
\label{fig13}
\end{figure}
A mode of
evolution was found in which the planetary migration was halted before
the protoplanet could approach the central binary and reach the 4:1
resonance. This only occurred
when the central binary
had an initial eccentricity of $e_{bin} \ge 0.2$. The
migration stalls because the circumbinary disc becomes eccentric.
We label this mode of evolution as `Mode 3'.
We illustrate this mode of evolution by presenting the results of a simulation
with $m_p=1$ Jupiter mass, $q_{bin}=0.1$, and $e_{bin}=0.2$.
Figure~\ref{fig13} shows snapshots of the surface density at different times
during the simulation, with the disc becoming noticeably eccentric.
Interaction between the protoplanet and the eccentric disc leads to a dramatic
reduction
or even reversal of the time--averaged torque driving the migration.
This is because the disc--planet interaction becomes dominated by the $m=1$
surface density perturbation in the disc rather than by the usual interaction
at Lindblad resonances in the disc.
\begin{figure}
\centerline{
\epsfig{file=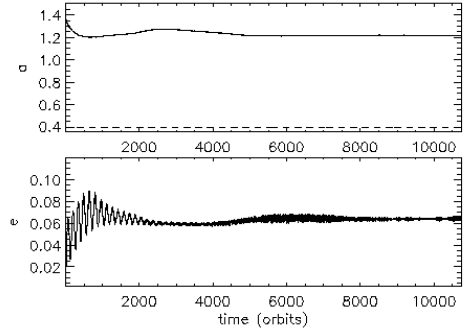,width=7cm} }
\caption[]
{This figure shows semimajor axis and eccentricity for Mode 3 run described
in text.}
\label{fig14}
\end{figure}
Figure~\ref{fig14} 
shows the evolution of the semimajor axis and eccentricity of the
planet, and shows that the migration stalls.
Simulations of this
type can be run for many thousands of planetary orbits without any significant
net inward migration occurring. Such systems
are likely to be stable long after the circumbinary disc has dispersed, since
the planets remain in the region of stability defined by the work of
Holman \& Wiegert (1999), and
are probably the best candidates for finding stable circumbinary extrasolar
planets. Interestingly, spectroscopic binary systems with significant
eccentricity are significantly more
numerous than those with lower eccentricities
(e.g. Duquennoy \& Mayor 1991; Mathieu et al. 2000), suggesting
that circumbinary planets may be common if planets are able to form in
circumbinary discs.

\section{Three Planet Resonant Systems}
\label{3planet}
The are a number of multiplanet systems among the known extrasolar
planets. In at least two, and possibly three, of these multiplanet systems,
two planets appear to be in mean motion resonance. The planets in the
systems GJ876 and HD82943 appear to be in 2:1 resonances, while two of
the planets in the 55 Cancri system appear to be in 3:1 resonance.
The existence of these mean motion resonances can be explained by a model
in which two planets form in the disc approximately coevally, and are
driven into resonance by differential inward migration due to interaction
with  the protostellar disc (e.g. Snellgrove, Papaloizou, \& Nelson 2001;
Lee \& Peale 2002; Nelson \& Papaloizou 2002). 

In this section we present some preliminary calculations that address 
the question of whether three--planet resonances may be formed, and
remain stable, by a process which is analogous to the above scenario when
each of the protoplanets is of Jovian mass.
Namely, approximately coeval formation of three planets which are
then driven into  a three--planet resonance through interaction with
the protoplanetary disc. We examine this question by means of 2--D
numerical simulations very similar to those described in section~\ref{binary}
using the same code and basic disc models. \\
We have considered a number of different scenarios, which we describe below:\\
({\it i}) Coeval formation of three planets, with the outer two forming a 2:1
resonance, followed by inward migration of the resonant pair
until the middle and inner planet themselves become locked in a 2:1 resonance.
We refer to this as a 4:2:1 resonance.\\
({\it ii}) Formation of two planets which differentially migrate and
lock into 2:1 resonance, followed by formation of a third planet at
larger radius that migrates in and locks into 2:1 resonance with the
middle planet. This is another example of a 4:2:1 resonance. \\
({\it iii}) A scenario very similar to scenario ({\it i}), except
the middle and inner planet form a 3:1 resonance. 
We call this a 6:2:1 resonance. \\
({\it iv}) A scenario similar to scenario ({\it ii}), except the
outer and middle planets form a 3:1 resonance. We call this a 6:3:1 resonance.

\subsection{4:2:1 Resonances}
\label{R4_2_1}
\begin{figure}
\centerline{
\epsfig{file=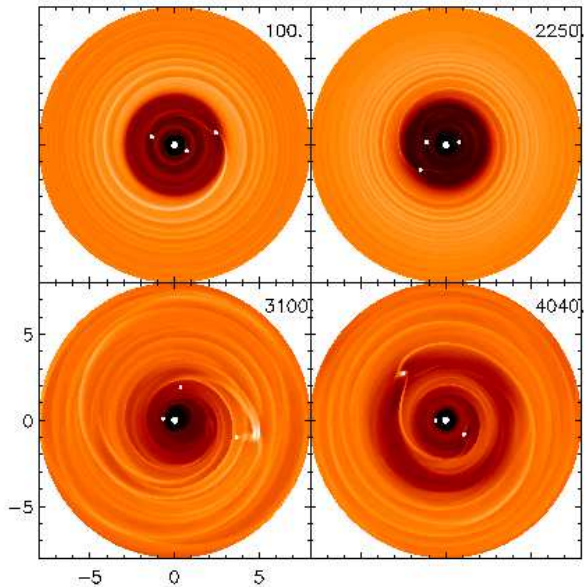,width=8cm} }
\caption[]
{This figure shows contour plots for the run designed to test scenario ({\it i})
described in text.}
\label{fig15}
\end{figure}
\begin{figure}
\centerline{
\epsfig{file=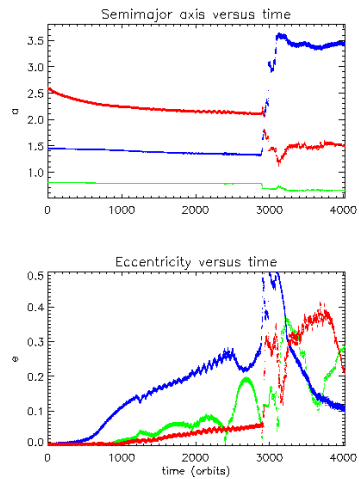,width=7cm} }
\caption[]
{This figure shows semimajor axes and eccentricities for three planets in run
designed to test scenario ({\it i}) described in text.}
\label{fig16}
\end{figure}

Both scenrios ({\it i}) and ({\it ii}) above could potentially lead
to the formation of a 4:2:1 resonance. In this subsection we present the results
of a simulation that examines scenario ({\it i}).
This simulation consisted of three Jovian mass protoplanets being inserted into
a protostellar disc model on fixed circular orbits
for a period of $\simeq 200$ orbits of the middle
planet until partial gap clearing had occurred. The protoplanets were
then allowed to evolve under the gravitational influence of the disc.
The evolution is shown in figure~\ref{fig15} which shows snapshots of
the evolution at four separate times. The orbital evolution is shown
in figure~\ref{fig16}. The outer planet migrates inward the most rapidly
and enters the 2:1 resonance with the middle planet. This resonant
pair then migrate in towards the innermost planet. As they approach
the orbital radius at which the middle planet becomes locked into 2:1 resonance
with the innermost planet, there is a strong gravitational interaction that
leads to scattering of the planets. Figure~\ref{fig16} shows
that the middle and outer planet interchange during this encounter
leading to the middle planet being flung out to larger radii in the disc.
All simulations that examine scenario ({\it i}) 
result in similar behaviour, with no 4:2:1 resonance ever forming.
However, interesting results are obtained which show that the kind of
interaction described here may help some planets survive migration in a disc
when they would otherwise migrate into the central star by being
ejected to larger radii. The evolution after ejection is likely to involve
gap formation and slow inward type II migration.

Simulations that test scenario ({\it ii}) suggest that 4:2:1 resonances
can be established, but that they are never stable for more than a few
hundred orbits This is primarily because the middle
planet has its eccentricity pumped up to large values
by virtue of undergoing resonant interaction with
both the inner most and outer most planets.
These will be described in a forthcoming paper
(Nelson \& Papaloizou 2003c).

\subsubsection{6:2:1 and 6:3:1 Resonances}

\begin{figure}
\centerline{
\epsfig{file=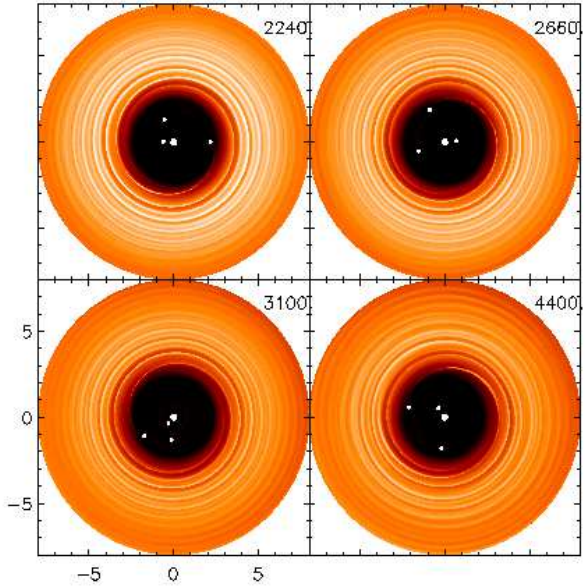,width=8cm} }
\caption[]
{This figure shows contour plots for run designed to examine 
scenario ({\it iii}) described in text.}
\label{fig17}
\end{figure}

\begin{figure}
\centerline{
\epsfig{file=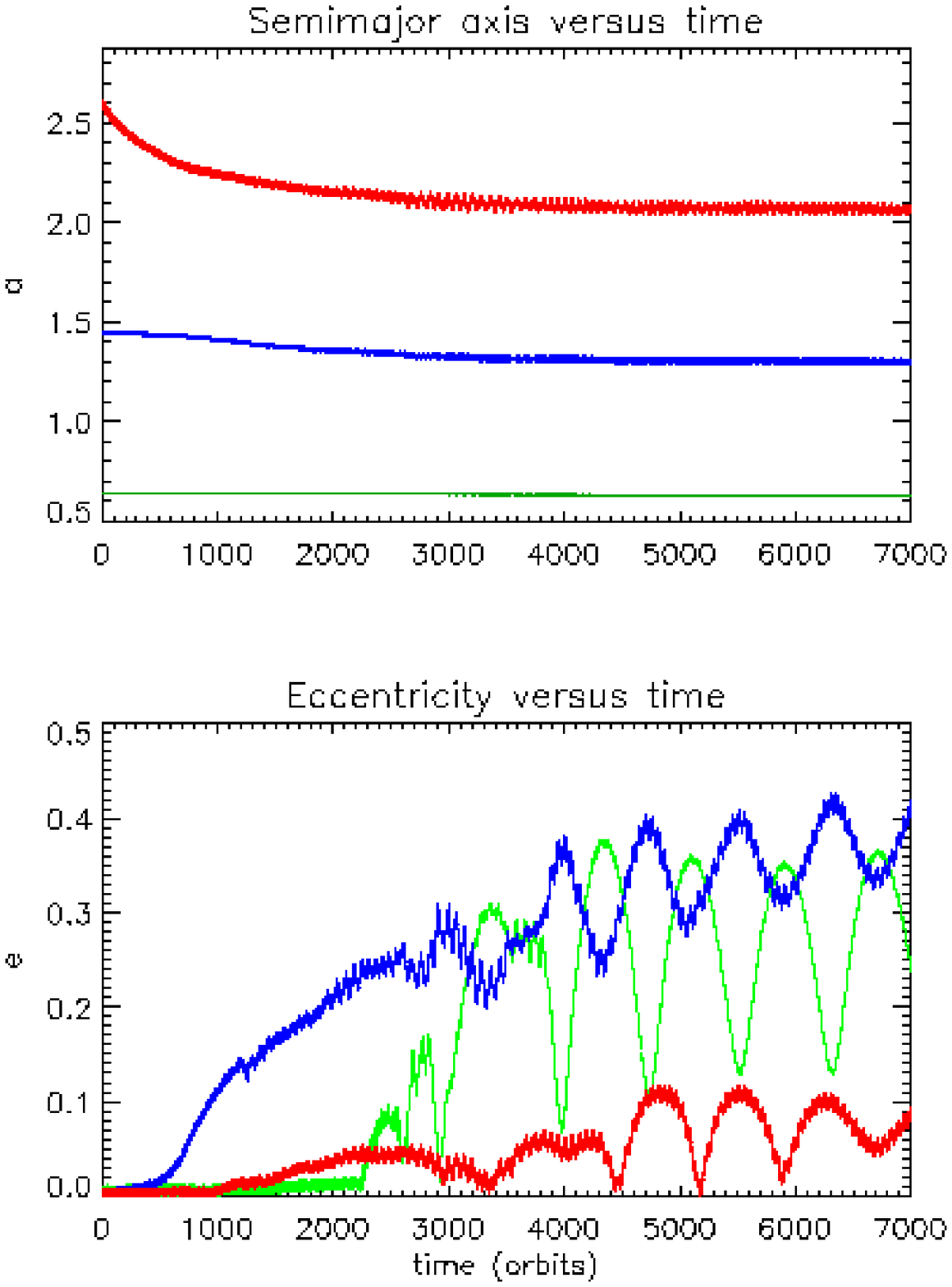,width=7cm} }
\caption[]
{This figure shows semimajor axes and eccentricities for three planets in run 
designed to examine scenario ({\it iii}) described in text.}
\label{fig18}
\end{figure}

\begin{figure}
\centerline{
\epsfig{file=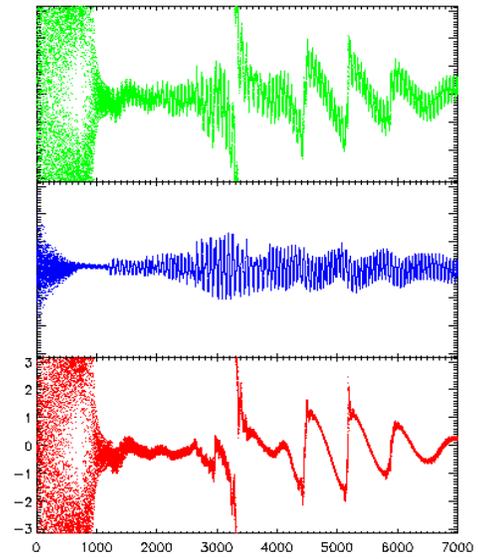,width=7cm} }
\caption[]
{This figure shows the resonant angles (upper two panels)
associated with 2:1 resonance for
outer two planets in scenario ({\it iii}) run described in text.
The lowest panel shows the difference in the longitudes of pericentre.}
\label{fig19}
\end{figure}

\begin{figure}
\centerline{
\epsfig{file=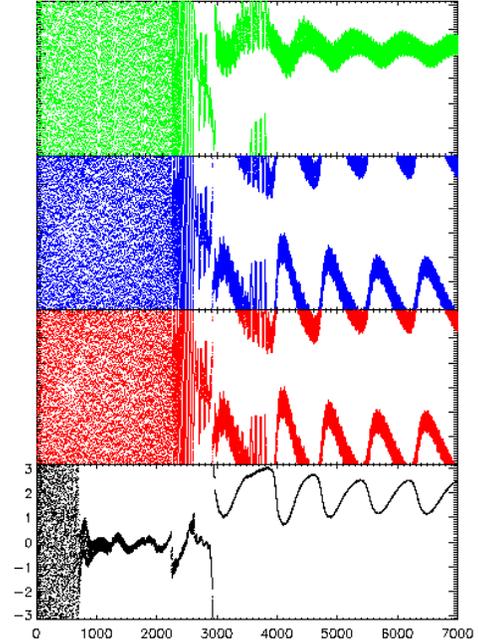,width=7cm} }
\caption[]
{This figure shows the resonant angles (upper three panels) associated with
3:1 resonance for middle and inner planet in scenario ({\it iii}) run
described in text. The lowest panel shows the difference in the longitude of
pericentre for the middle and inner planet in this run.}
\label{fig20}
\end{figure}

Simulations have been performed to examine the formation of 6:2:1
and 6:3:1 resonances. These indicate that such resonances may be formed
quite easily, and remain stable over long time periods.
In this section we present one simulation that examined the plausibility
of scenario ({\it iii}). \\
The initial setup was similar to that described in section~\ref{R4_2_1}
except that the inner two planets were more widely separated.
The outer most planet migrated inward the most rapidly and locked into
2:1 resonance with the middle planet. The resonant pair then migrated in towards
the inner most planet until the middle and inner planet became locked into
3:1 resonance. Figure~\ref{fig17} shows snapshots of the disc and planet 
evolution. The orbital evolution is shown in figure~\ref{fig18}, which 
shows the converging orbits, and the resulting slow-down in migration
as the three planets become resonantly locked leading to an effective
increase in the inertia of the system that is being driven inward
by the disc. The resonant angles for the outer two planets 
are shown in figure~\ref{fig19}, showing the establishment of the
2:1 resonance, and the resonance angles for the inner two planets
are shown in figure~\ref{fig20} showing the establishment of the 3:1 resonance.
Also shown in the lowest panels of figures~\ref{fig19}) and \ref{fig20}
are the differences in the longitudes of pericentre for each of the planet pair.
Figure~\ref{fig20}
shows that the middle and inner planet becomes apsidally locked 
when the outer two planets enter the 2:1 resonance, since the eccentricity
driving by the resonance acts as an effective `tuning' mechanism for
the establishment of a secular resonance between the inner two planets.
We note that such a secular resonance is observed in the 
Upsilon Andromeda system.

\section{Summary}
\label{summary}
In this article we have presented the results from three distinct projects.
The first examined the interaction between embedded protoplanets and
turbulent, magnetised discs. Broadly speaking we find rather
similar behaviour for high mass, gap forming planets when compared
with previous work on laminar but viscous discs. However, low mass
protoplanets behave rather differently, and appear to under go 
`stochastic migration' due to interaction with the background turbulence.
The result appears to be that low mass protoplanets will undergo a randon
walk in turbulent discs, instead of monotonic inward migration. \\
We presented the results of simulations that examined the formation of
giant protoplanets in circumbinary discs. We find that under a wide range of
conditions stable circumbinary planets may be maintained. In particular,
if the binary system is significantly eccentric, then disc induced inward
migration of the protoplanet does not occur. \\
Finally, we presented some preliminary calculations of three protoplanets
forming in a disc, and examined the conditions under which three--planet
resonances could be established and maintained. We found that 4:2:1
configurations could be formed, but quickly became unstable.
However, 6:2:1 and 6:3:1 configurations could form and remain stable over
very long periods.

\begin{center}
{\large ACKNOWLEDGMENTS} 
\end{center}
The computations reported here were performed using the 
UK Astrophysical Fluids Facility (UKAFF)

\end{document}